\documentclass[a4paper]{jpconf}

\usepackage{url}
\usepackage{hyphenat}
\usepackage{microtype}
\usepackage{hyperref}
\usepackage{graphicx}
\usepackage{subfigure}
\graphicspath{{figures/}}

\usepackage{silence} 

\usepackage{hyperref}
\hypersetup{
  colorlinks, breaklinks, linkcolor=black, urlcolor=black, citecolor=black,
  pdfauthor  = {G Amadio, J Apostolakis, P Buncic, G Cosmo, D Dosaru, A Gheata, S Hageboeck, J Hahnfeld, M Hodgkinson, B Morgan, M Novak, A A Petre, W Pokorski, A Ribon, G A Stewart and P M Vila},
  pdftitle   = {Offloading electromagnetic shower transport to GPUs}
}

\begin{document}

\title{Offloading electromagnetic shower transport to GPUs}

\author{G~Amadio$^1$, J~Apostolakis$^1$, P~Buncic$^1$, G~Cosmo$^1$, D~Dosaru$^2$, A~Gheata$^1$, S~Hageboeck$^1$, J~Hahnfeld$^1$, M~Hodgkinson$^3$, B~Morgan$^4$, M~Novak$^1$, A~A~Petre$^{5,6}$, W~Pokorski$^1$, A~Ribon$^1$, G~A~Stewart$^1$ and P~M~Vila$^1$}

\address{$^1$ CERN, Geneva (CH)}
\address{$^2$ EPFL, Lausanne (CH)}
\address{$^3$ University of Sheffield (GB)}
\address{$^4$ University of Warwick (GB)}
\address{$^5$ University of Bucharest (RO)}
\address{$^6$ University Politehnica of Bucharest (RO)}

\ead{andrei.gheata@cern.ch}

\begin{abstract}
Making general particle transport simulation for high-energy physics (HEP) single-instruction-multiple-thread (SIMT) friendly, to take advantage of accelerator hardware, is an important alternative for boosting the throughput of simulation applications. To date, this challenge is not yet resolved, due to difficulties in mapping the complexity of Geant4 components and workflow to the massive parallelism features exposed by graphics processing units (GPU). The AdePT project is one of the R\&D initiatives tackling this limitation and exploring GPUs as potential accelerators for offloading some part of the CPU simulation workload. Our main target is to implement a complete electromagnetic shower demonstrator working on the GPU. The project is the first to create a full prototype of a realistic electron, positron, and gamma electromagnetic shower simulation on GPU, implemented as either a standalone application or as an extension of the standard Geant4 CPU workflow. Our prototype currently provides a platform to explore many optimisations and different approaches. We present the most recent results and initial conclusions of our work, using both a standalone GPU performance analysis and a first implementation of a hybrid workflow based on Geant4 on the CPU and AdePT on the GPU. 
\end{abstract}

\section{Introduction}
\label{sec:introduction}
Several online and offline applications in high-energy physics (HEP) have benefited from running on graphics processing units (GPUs), taking advantage of their processing model (\cite{Aaij2020,aliceGPU,10.3389/fdata.2020.601728} are some of the more notable examples). To date, however, general HEP particle transport simulation is not one of them, due to difficulties in mapping the complexity of its components and workflow to the GPU’s hardware. Deep polymorphic code stacks, low branch predictability, incoherent memory accesses, and the use of stateful global managers are significant obstacles that prevent porting Geant4~\cite{Agostinelli:2002hh} to GPUs. However, HEP computing will need to exploit more and more heterogeneous resources in the future \cite{Albrecht2019}, and our current inability to use GPU cards, which are increasingly found in computing centres, for detailed simulation of collider experiments limits our ability to exploit hardware that is available to HEP and therefore raises costs, which motivates R\&D in this area.

The AdePT~\cite{AdePT} project started late 2020 with the goal of demonstrating a complete workflow working on GPUs, having all the simulation stepping components: complete physics models describing the electromagnetic (EM) processes, magnetic field propagation in detector geometry, and code producing user hit information transferred from the GPU back to the host. Existing CPU simulation components had to be adapted, extended, or redesigned entirely to fit to the GPU's hardware. The project evolved from very simple examples, which demonstrated only limited functionality. Geometry was one of the first usable modules, and was provided by the VecGeom~\cite{Apostolakis:2015:ACAT} library. The G4HepEm library~\cite{G4HepEm} was utilised for modelling EM interactions of electron, positron and gamma particles on the device. Finally, we developed a generic example demonstrating the integration of Geant4 CPU-based simulation with the AdePT GPU workflow.  

The paper describes our most recent validation and performance results for setups with different complexity. We present our initial conclusions and our preliminary understanding of the usability of GPUs for full simulation in HEP applications.

\section{Simulation components}
\label{sec:simulation_components}
The AdePT workflow was built on top of the GPU ports of the main particle transport simulation components. While the VecGeom GPU functionality pre-dated our project, modules such as physics and magnetic field were rewritten or adapted from their original CPU-based versions.

\subsection{Geometry integration, validation and optimisation}
Geometry was the first major simulation component adapted to the AdePT GPU workflow, 
building on existing CUDA support in the VecGeom library. 
VecGeom's existing capabilities included generating and compiling CUDA code for all geometry types 
(solids, placed and unplaced volumes) in a namespace separate from CPU/host code, and copying 
the CPU transient model to the device.  
Code to navigate on a GPU was not available.

We have made an early 
implementation of a ray-tracing demonstrator using the same executable on host and device, and 
a common geometry service usable by other AdePT examples. 
Among the downsides of the current VecGeom GPU port, direct binding to CUDA is required to support virtual polymorphism, in particular for different solid types. This currently prevents the usage of portability frameworks and other GPU backends.

First a simple 'loop' navigator was created, which searches all daughter volumes, to serve as a baseline to 
validate correctness of other approaches.
Development of the navigation capabilities concentrated on creating acceleration structures that can
be used on the GPU.
A bounding volume hierarchy (BVH) per logical volume was implemented in GPU-portable 
data structures and code, to reduce the number of candidate volumes that must be intersected. We observed performance improvements of up to 2.4x in case of a CMS geometry setup from 2018 compared to the 'loop' navigator case.
Since geometry navigation was identified as one of the major bottlenecks in the GPU workflows, 
additional optimisations were undertaken:
\begin{itemize}
  \item a separate kernel was used to manage volume relocation, to improve load balancing;
  \item a revision of boundary-crossing eliminated a bias and reduced the number of trial steps needed;
  \item the storage of the navigation hierarchy information in navigation states was improved, reducing its memory footprint;
  \item we implemented the option to compile all types and interfaces of VecGeom in single-precision. Using ray-tracing in single-precision for the TrackML challenge \cite{Golling:2018yyi} geometry setup, we observed 44\,\% better performance on GPU, but also 14\,\% improvement on the CPU.
\end{itemize}
Together these resulted in important consistency and performance improvements compared to the initial non-optimised baseline. 

\begin{figure}[ht]
\hspace*{3pc}\includegraphics[width=0.3\textwidth]{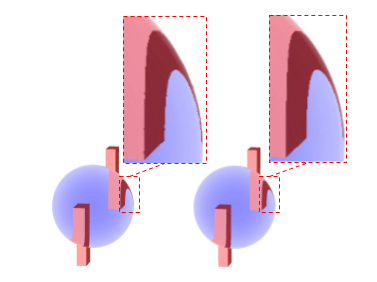}\hspace{6pc}%
    \begin{minipage}[b]{0.45\textwidth}
        \caption{Geometry validation using ray tracing, comparing single-precision (left) with the default double-precision (right) images.}
        \label{fig:adept-raytrace-sd}
    \end{minipage}
\end{figure}

Geometry navigation validation was initially done using images from a ray-tracing example, comparing at pixel level images produced for setups of different complexity on GPU against results obtained with the same executable running on CPU.
This setup allowed the validation of other geometry optimisations, such as single-precision support in navigation, as shown in figure~\ref{fig:adept-raytrace-sd}. Geometry navigation was also validated using more advanced examples embedding physics-driven stepping, by comparing step-length distributions, and other low-level observables, with the values obtained in Geant4 reference setups.

\label{sec:geometry}

\subsection{Physics integration and validation}
AdePT uses the G4HepEm library~\cite{G4HepEm} to model the electromagnetic physics of electrons, positrons, and gammas.
The library is a compact rewrite of the three particles' electromagnetic processes that are relevant for HEP detector simulation.
It follows a strict separation of initialisation, that is dependent on Geant4, and run-time functionalities, which are standalone.
G4HepEm also directly interfaces with CUDA to transfer the required physics data to GPUs.

For the AdePT project, we introduced two generalisations into the run-time functions.
Firstly, we abstracted the usage of the random number generator (RNG).
On the host, and when integrating with Geant4, G4HepEm uses the RNGs provided by CLHEP.
However, we choose a portable implementation of RANLUX++ on the GPU which was described recently~\cite{RANLUX++}.
Secondly, it was necessary to split some functions that accessed G4HepEm's thread-local storage.
Accepting arguments as scalar values allows utilisation of track storage that is more appropriate for parallel processing on the GPU.
With these two changes, it is possible to reuse 95\,\% of G4HepEm and only write some glue code in AdePT itself.
This greatly reduced the time and effort needed for this part of the demonstrators.

\begin{figure}[ht]
\includegraphics[width=0.5\textwidth]{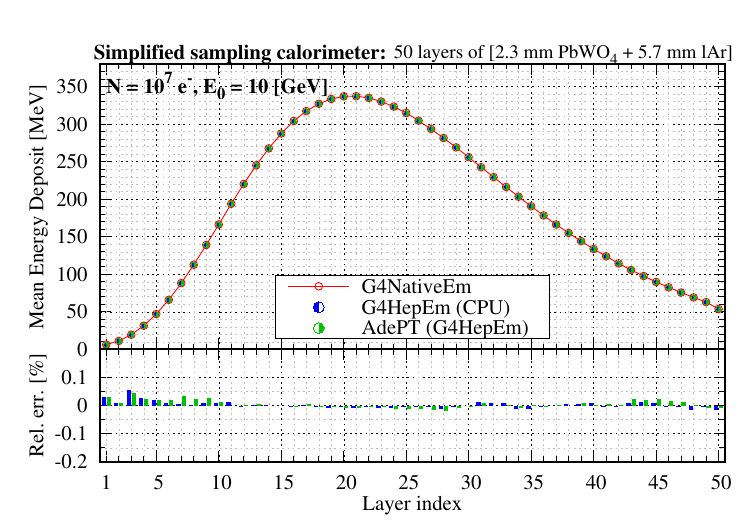}\hspace{2pc}%
    \begin{minipage}[b]{0.43\textwidth}
        \caption{Mean energy deposit in a simplified sampling calorimeter for Geant4, G4HepEm on the CPU, and AdePT.}
        \label{fig:validation}
    \end{minipage}
\end{figure}

For validating the physics integration, we simulate the mean energy deposit in a simplified sampling calorimeter.
The calorimeter consists of 50~layers, each made up of slabs of PbWO$_4$~(2.3\,mm) and liquid argon~(5.7\,mm).
The choice of lead tungstate was made to test the element selectors for a material composed of multiple elements.
The same setup is also used for the validation of G4HepEm on the CPU against Geant4.
Thus, figure~\ref{fig:validation} shows a three-fold comparison with excellent agreement between Geant4, G4HepEm on the CPU, and AdePT.
We further compared the number of secondary particles generated by G4HepEm on the CPU with AdePT, and found these to be equal to per-mil level.

\label{sec:physics}

\subsection{Magnetic field}
A capability to propagate charged tracks in a constant magnetic field has been incorporated into the prototype.  
In order to intersect the model geometry, the track is broken into linear segments.
In each segment the estimated deflection of the curved path from the projection
of the direction must be less than or equal to an adjustable threshold.
Breaking up a step into multiple linear track segments can create significant divergence
between threads.  In addition, refining a candidate intersection to its precise location
is a secondary source of divergence.

An investigation over the effects of the divergence of threads used a maximum number 
of iterations. Tracks which did not finish their integration were killed and accounted for, in 
order to measure their overall contribution to the total energy deposited.
A large value for the maximum number of iterations (1000) was required to 
ensure that less than $0.5\%$ of the energy is lost in a geometry from a LHC detector.
Even at the smaller value like $100$, which corresponds to $3\%$ of energy deposit,
the overall simulation time was about $4$ times slower than when using a maximum value 
of $10$ ($8\%$ energy deposit).
These figures demonstrate how vital it is to handle this divergence.  
Approaches to address it are being evaluated.

%
%
\label{sec:field}

\section{Stepping workflow}
In Geant4, every thread works on a single particle track, transporting it one step at a time.
Secondary particles are pushed onto a stack, and consumed as threads finish processing their current particles.
To make use of multiple cores, events are distributed across threads, which continue to work independently of each other.
While efficient on the CPU, this approach does not yield enough uniformity and parallelism for GPUs.
Instead, the demonstrators developed in the AdePT project immediately start tracking secondary particles.
Parallelism is achieved by making one step for an entire population of tracks.
This requires creating secondaries in parallel, without influencing the current population of tracks.

In the current demonstrators, we implement this with arrays that all threads can append to in parallel.
For each particle type (electron, positron, gamma), there are two arrays that represent the current and the next population, respectively.
They contain indices into the track storage, and are swapped after each step.
Having distinct arrays per particle type has the advantage that it is not necessary to store the type explicitly.
Rather, it is implicit from the array that the track is taken from or inserted into.
To remain efficient, we launch the three kernels for transporting all particles of each type into different CUDA streams.
This allows the hardware to run them concurrently, after which we synchronise using CUDA events.

At the time of writing, track storage is implemented as a contiguous array with an atomic counter.
While simulating a batch of particles, the kernels only append to the storage.
When a track is finished, its slot is not reused which is wasteful in terms of memory.
However, any more elaborate solution will require more synchronisation and introduce overhead.
For this reason, we decided to postpone optimisation in this area until memory usage becomes an issue.
\label{sec:workflow}

\section{Geant4 integration}
Besides the focus on functionality and performance, the integration with the Geant4 CPU workflow is very important to explore, given that hadronic physics is not yet implemented on the GPU. We based our first integration implementation on the existing Geant4 fast-simulation hooks. Those hooks consist of defining a `fast simulation' process responsible for a region of the geometry where, for particles passing some selection criteria (particle type, energy, etc), an external simulation library is called. Those particles are taken off the Geant4 stack, and the handling of the detector response is delegated to the `fast simulation' process. 


The AdePT library can indeed be considered as such `fast simulation' process, which takes particles from Geant4 as input and gives back the energy depositions and any 'outgoing' particles from that region as output. The upside of such an approach is that it allows configurable activation of the AdePT integration and easy comparison with native Geant4 simulation using a single executable. 

The current GPU workflow of AdePT was proven to be more efficient with increasing the input tracks batch size. This was implemented in the current integration by buffering particles entering the 'GPU region'. Buffering is done until reaching a configurable threshold, which triggers synchronous processing of the batch on the GPU. This approach required adding a new \textit{Flush} method in the fast-simulation process interface, called within the main Geant4 event loop, which enables processing of any remaining particles in the buffer when the Geant4 event stack is empty. 

Once the AdePT batched processing completes, the particles outgoing from the GPU-enabled region are passed back onto the Geant4 stack. The \textit{Flush} interface is called in the event loop when the Geant4 'urgent stack' is empty, but before the event is finished. If particles are added back onto the Geant4 stack, the Geant4 event will continue to process those. This interface is usable by any `fast simulation' models which rely on buffering input particles to process them concurrently.

As far as the energy depositions (hits) produced by AdePT are concerned, they are stored in arrays transferred back from the device to the host at the end of each event. The corresponding Geant4 hits are then created (assigned to specific sensitive volumes) and added to the Geant4 hit collections. This allows for structurally identical simulation output when running either with Geant4 only or with Geant4 plus AdePT. 

\label{sec:integration}

\section{Preliminary analysis}
\label{sec:analysis}
\begin{figure}[ht]
    \centering
    \includegraphics[width=\textwidth]{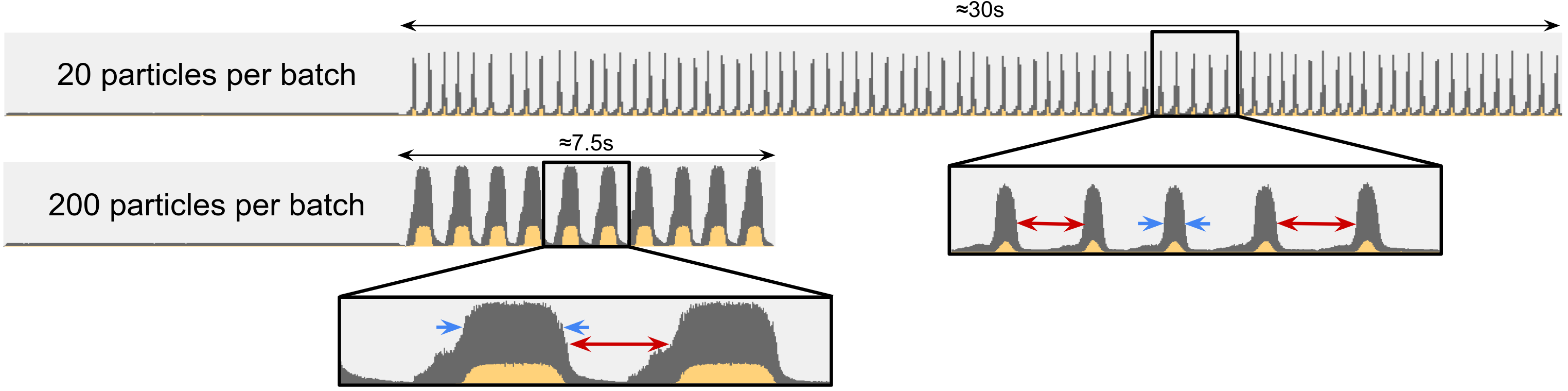}
    \caption{GPU usage for two runs of an AdePT simulation of CMS on an Nvidia RTX 2070 GPU, one with 20 particles per batch, and another with 200 particles per batch. Each run simulates 2000 primary particles. Colors represent unallocated warps in active SMs (gray), and compute warps in flight (yellow).}
    \label{fig:adept-gpu-usage}
\end{figure}

A preliminary performance analysis of AdePT has been done, which provides some insights into how further progress can be made in this area.
In figure~\ref{fig:adept-gpu-usage}, the GPU usage for two runs with different particle batch sizes is shown.
As the electromagnetic shower develops for each batch of particles, GPU usage increases.
When the average energy of the particles is no longer enough to create new secondaries, GPU usage begins to decrease.
Hence each peak in usage corresponds to one batch of particles being simulated.
Small batches lead to sharp peaks with significant amounts of idling time between them, while larger batches make for better use of the GPU resources with broader peaks and about 4x faster run time.
Nevertheless, even with large batches of particles, there is still some idling time between the peaks that could be reduced by injecting new primary particles into the simulation at regular intervals rather than simulating full batches until the end.
This would allow for staggered showers to develop for each set of injected particles, increasing the overlap and reducing the low GPU usage intervals between the peaks currently observed.

\begin{figure}[h]
    \centering
    \subfigure[20 particles]{\includegraphics[height=0.11\textwidth]{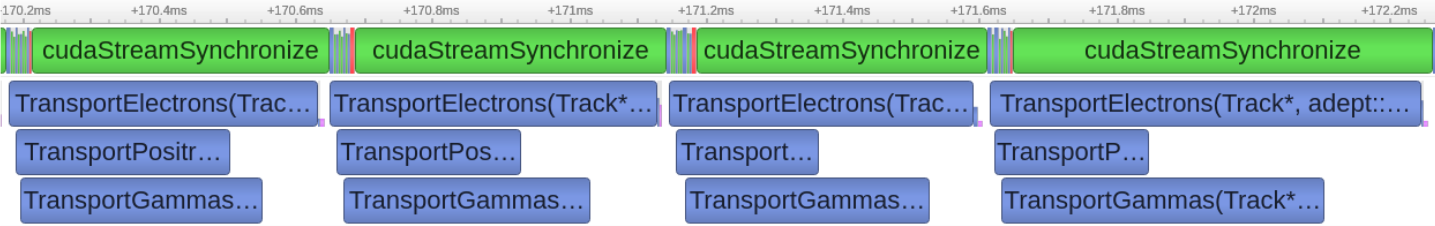}}
    \subfigure[200 particles]{\includegraphics[height=0.11\textwidth]{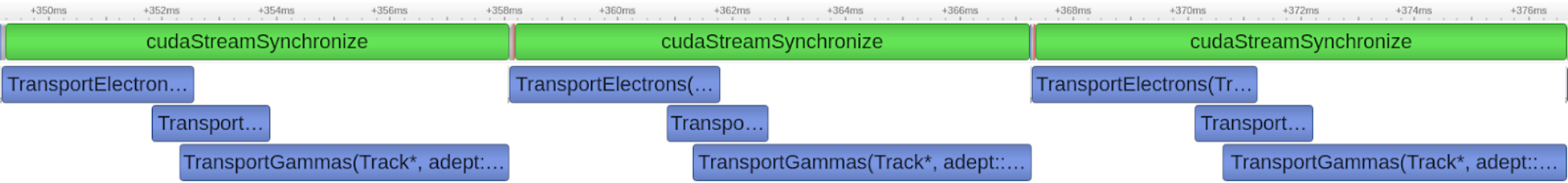}}
    \caption{Detailed view of a few simulation steps and their kernels at peak GPU usage.}
    \label{fig:adept-kernels}
\end{figure}

Figure \ref{fig:adept-kernels} provides a closer look at a few simulation steps when GPU usage is at its peak.
On one hand, in small batches, there are not enough warps in flight to fill the GPU, which means that all kernels can start running at the same time, and the slowest kernel determines the pace of the simulation.
The white space between kernel calls represents idle time on the GPU in this case.
On the other hand, in large batches, a single kernel has enough warps to fill the whole GPU, so other kernels can only start after some of the warps from the previous kernel have finished executing.
In this case, the white space at the beginning of each step is an indication that the GPU is busy, and idle time is limited to the end of the step, when the last running kernel no longer has enough warps to fill the whole GPU.

Complementary to analysing the overall device occupancy, we analysed the performance of a single geometry navigation kernel on a Tesla V100 data-centre GPU at peak occupancy.
While this device can theoretically handle 16 warps per scheduler, the register usage for handling rotation matrices and deep call stacks limits the kernel to only three warps per scheduler.
Of these three warps, an average of 0.1 warps are eligible for execution at a given time.
The device was found to issue on average one instruction every 34 cycles.
Two main causes for the low in-kernel utilisation were identified.
Firstly, when many particles navigate the on-device detector description, quasi-random memory access is required.
Given that GPUs are not optimised for low memory latency, warps regularly stall when waiting for data from device memory.
Secondly, VecGeom uses virtual functions to compute distances to geometric objects.
If the threads within a warp jump into different functions, the warp diverges, effectively serialising threads within the warp.
It is likely possible to speed up these computations, but this requires a redesign of the on-device detector description.

\section{Conclusions and next steps}
\label{sec:conclusions}
The AdePT project implemented a first full prototype of particle transport on GPUs, having all the simulation stepping components: complete physics models describing the electromagnetic processes, magnetic field propagation in detector geometry, and code producing user hits data transferred from the GPU back to the host. The main simulation components were re-engineered to be GPU aware, as either GPU functional ports in the case of geometry, or complete re-implementations in the case of the physics library.

Besides the required functionality, performance portability was one of the aspects investigated. We have developed code abstractions based on portability libraries such as Alpaka ~\cite{ZenkerAsHES2016} and OneApi~\cite{oneapi}, but also developed a kernel launcher abstracted on the backend type. It was found that portability can only be achieved if more effort is invested, due to both the maturity of the available tools, and the fact that VecGeom GPU support is only implemented for CUDA. A revisited geometry GPU port would allow resuming the portability work at a future stage.

Performance is one of the important aspects we monitored upon implementing more complex examples. This triggered specific performance optimisation developments, from handling random numbers to geometry navigation and host-device interplay in the hybrid simulation approach. Geometry was identified as being the main bottleneck in the current project phase, limiting the device usage to a small fraction of the theoretical maximum. This inefficiency became irreducible after doing a number of obvious optimisations, being intrinsic to the geometry model used, deemed too complex and unbalanced for the GPU. This puts the overall GPU efficiency target out of reach in the current implementation, and calls for a deeper review of the adopted GPU approaches for geometry and other work-divergent components, such as magnetic field propagation.

Having implemented a hybrid workflow prototype able to use the GPU as opportunistic accelerator in a general Geant4-driven CPU application represents a major milestone for the AdePT project. Optimisation work for the application has just started, aiming to expose additional opportunities for performance improvement. The AdePT state-of-the-art prototype provides a platform for understanding both the opportunities and future efforts needed to allow general HEP detector simulation to benefit from GPU accelerator hardware.


\ack
The oneAPI related work was conducted with the support of Intel in the framework of the CERN openlab-Intel collaboration agreement. 
The authors acknowledge the funding from EPSRC and STFC agencies.
The single-precision work was supported by the GSoC 2021 program.

\section*{References}
\bibliographystyle{iopart-num}
\bibliography{AdePT_ACAT21_paper}

\end{document}